\documentclass[conference]{IEEEtran}
\IEEEoverridecommandlockouts

\makeatletter
\def\markboth#1#2{\def\leftmark{\@IEEEcompsoconly{\sffamily}\MakeUppercase{\protect#1}}%
\def\rightmark{\@IEEEcompsoconly{\sffamily}\MakeUppercase{\protect#2}}}
\makeatother

\usepackage[english]{babel}
\usepackage{color}
\usepackage{lipsum}
\usepackage{caption}
\usepackage[noadjust]{cite}
\usepackage[pdftex]{graphicx}
\usepackage{caption}
\usepackage{amsmath}
\usepackage{amsfonts}
\usepackage{amssymb}
\usepackage{amsthm}
\usepackage{array}
\usepackage{verbatim}
\usepackage{listings}
\usepackage{algorithm}
\usepackage{algpseudocode}
\usepackage{url}
\usepackage{enumerate}
\usepackage{multirow}

\usepackage{epsfig}
\usepackage{epstopdf}
\usepackage{multicol}
\usepackage[font=footnotesize]{caption}
\usepackage[font=footnotesize]{subcaption}

\usepackage{makecell}
\usepackage{tablefootnote}

\renewcommand{\arraystretch}{2}

\newcommand{\bi}{\begin{itemize}}
\newcommand{\ei}{\end{itemize}}
\newcommand{\be}{\begin{equation}}
\newcommand{\ee}{\end{equation}}

\def\beq{\begin{equation}}
\def\eeq{\end{equation}}
\def\beqa{\begin{eqnarray}}
\def\eeqa{\end{eqnarray}}
\def\beqan{\begin{eqnarray*}}
\def\eeqan{\end{eqnarray*}}

\usepackage{threeparttable}
\usepackage[table,xcdraw]{xcolor}
\usepackage{tabularx}
   \usepackage{multirow}
   \usepackage{booktabs}
   \usepackage{graphicx}
   \usepackage{array, blindtext}
   \usepackage{wrapfig}
\usepackage{pdfpages}
\usepackage{xfrac}
\usepackage{tikz}
\usepackage{pgfplots}
\pgfplotsset{compat=newest} 
\pgfplotsset{plot coordinates/math parser=false} 
\newlength\fheight
\newlength\fwidth
\usetikzlibrary{plotmarks,patterns,decorations.pathreplacing,backgrounds,calc,arrows,arrows.meta}
  \usepgfplotslibrary{patchplots}
  \usepackage{grffile}
  \usepackage{amsmath}
\usepackage[acronym]{glossaries}
\usepackage{adjustbox}

\title{Will TCP work in mmWave 5G Cellular Networks?\vspace{-3mm}}

\author{Menglei Zhang$^\dagger$, Michele Polese$^*$, Marco Mezzavilla$^\dagger$,\\Jing Zhu$^\diamond$, Sundeep Rangan$^\dagger$, Shivendra Panwar$^\dagger$, Michele Zorzi$^*$\\
\small $^\dagger$NYU Wireless, New York University, NY, USA - 
e-mail: \{menglei, mezzavilla, srangan, panwar\}@nyu.edu\\
 \small $^*$Department of Information Engineering, University of Padova, Italy - 
e-mail: \{polesemi, zorzi\}@dei.unipd.it\\
$^\diamond$Intel Corporation - e-mail: jing.z.zhu@intel.com}

\newacronym[plural=MMEs,firstplural=Mobility Management Entities (MMEs)]{mme}{MME}{Mobility Management Entity}
\newacronym{ran}{RAN}{Radio Access Network}
\newacronym{cn}{CN}{Core Network}
\newacronym{m2m}{M2M}{Machine to Machine}
\newacronym{nfv}{NFV}{Network Function Virtualization}
\newacronym{vm}{VM}{Virtual Machine}
\newacronym{son}{SON}{Self-Organizing Network}
\newacronym{iot}{IoT}{Internet of Things}
\newacronym{rss}{RSS}{Received Signal Strength}
\newacronym{ttt}{TTT}{Time-to-Trigger}
\newacronym{enb}{eNB}{evolved Node Base}
\newacronym{scoot}{SCOOT}{Split Cycle Offset Optimization Technique}
\newacronym{utc}{UTC}{Urban Traffic Control}
\newacronym{tfl}{TfL}{Transport for London}
\newacronym{ue}{UE}{User Equipment}
\newacronym{hetnet}{HetNet}{Heterogeneous Network}
\newacronym{snr}{SNR}{Signal to Noise Ratio}
\newacronym{sinr}{SINR}{Signal to Interference plus Noise Ratio}
\newacronym{los}{LOS}{Line of Sight}
\newacronym{nlos}{NLOS}{Non-Line of Sight}
\newacronym{mec}{MEC}{Mobile Edge Cloud}
\newacronym{cc}{CC}{Congestion Control}
\newacronym{aqm}{AQM}{Active Queue Management}
\newacronym{ecn}{ECN}{Explicit Congestion Notification}
\newacronym{sack}{SACK}{Selective Acknowledgment}
\newacronym{aimd}{AIMD}{Additive Increase Multiplicative Decrease}
\newacronym{bdp}{BDP}{Bandwidth-Delay Product}
\newacronym{rtt}{RTT}{Round Trip Time}
\newacronym{rlc}{RLC}{Radio Link Control}
\newacronym{mss}{MSS}{Maximum Segment Size}
\newacronym{mtu}{MTU}{Maximum Transmission Unit}
\newacronym{rr}{RR}{Round Robin}
\newacronym{cdf}{CDF}{Cumulative Distribution Function}
\newacronym{harq}{HARQ}{Hybrid Automatic Repeat reQuest}
\newacronym{gnb}{gNB}{Next Generation Node Base}
\newacronym{quic}{QUIC}{Quick UDP Internet Connection}
\newacronym{lte}{LTE}{Long-Term Evolution}
\newacronym{codel}{CoDel}{Controlled Delay Management}
\begin{document}

\makeatletter
\patchcmd{\@maketitle}
  {\addvspace{0.5\baselineskip}\egroup}
  {\addvspace{0\baselineskip}\egroup}
  {}
  {}
\makeatother

\maketitle

\begin{abstract}
The vast available spectrum in the
millimeter wave (mmWave) bands offers the possibility of multi-Gbps data rates for fifth generation (5G) cellular networks. However, 
mmWave capacity can be highly intermittent due to the 
vulnerability of mmWave signals to blockages
and delays in directional searching.
Such highly variable links
present unique challenges for adaptive control mechanisms in 
transport layer protocols
and end-to-end applications.
This paper considers the fundamental question of whether TCP
-- the most widely used transport protocol -- will work
in mmWave cellular systems.
The paper provides a comprehensive simulation study of TCP 
considering various factors such as the congestion control algorithm, including the recently proposed TCP BBR,
edge vs.\ remote servers, handover and multi-connectivity, TCP packet
size and 3GPP-stack parameters. We show that the performance of TCP on mmWave links is highly dependent on different combinations of these parameters, and identify 
the open challenges in this area.

\end{abstract}

\begin{picture}(0,0)(0,-360)
\put(0,0){
\put(0,-10){\footnotesize This paper has been accepted for publication in} 
\put(0,-20){\footnotesize the IEEE Communication Magazine}}
\end{picture}

\flushbottom



\section{Introduction}
End-to-end connectivity over the internet largely relies on transport protocols that operate above the network layer. 
The most widely used transport protocol is the Transmission Control Protocol (TCP), designed in the 1980s~\cite{Postel:81} to offer reliable packet delivery and sending rate control to prevent congestion in the network. 
Reliability is accomplished with receiver's acknowledgments (ACKs) fed back to the sender, which retransmits packets if needed, while
rate control is achieved by dynamically adjusting the congestion window, i.e., the maximum amount of data that the sender can transmit without receiving ACKs. Several \gls{cc} algorithms have been proposed in order to improve the goodput (defined as the application layer throughput) and latency of TCP over different types of networks~\cite{liu2016improving}.

However, the next generation of cellular networks will present new challenges for TCP\footnote{In this work, we focus on TCP since it is the dominant transport protocol in use today. One possible avenue for future work is to consider the UDP protocol, that, however, shifts the burden of retransmissions and flow control to a higher layer, introducing similar problems as those related to TCP.}, specifically related to mmWave links in the radio access network, which exhibit an erratic propagation behavior. This technology is seen as a promising enabler for the 5G targets of multi-gigabit/s data rates and ultra-low latency~\cite{xiao2017millimeter}, but the end-to-end performance 
perceived by the user will eventually depend on the interaction with transport protocols such as TCP. Some recent studies~\cite{zhang2016transport,polese2017tcp} have highlighted that the extreme variability of the signal quality over mmWave links yields either a degraded TCP goodput and a very low utilization of the resources 
at mmWave frequencies, or, in the presence of link-layer retransmissions, high goodput at the price of high latency. 
Moreover, in~\cite{zhang2016transport} it is shown that the bufferbloat phenomenon (i.e., the increase in latency that is caused by excessive buffering) emerges as a consequence of the presence of large buffers in the network.

Our goal is to answer the question: \emph{Will TCP work in mmWave 5G cellular networks?} To this aim, we compare the performance of different TCP congestion control algorithms over simulated 5G end-to-end mmWave networks considering (1) a high speed train and (2) an urban macro 3GPP deployment, as further described in Sec.~\ref{sec:scenarios}. 
Our detailed simulation study demonstrates that the performance of TCP over mmWave depends critically on several aspects of the~network:

\begin{enumerate}
\item \textbf{Edge vs. Remote Server:} By comparing the end-to-end performance at varying server's location, we show that for a shorter control loop, i.e., when the server is placed at the cellular network edge, TCP can react faster to link impairments. 
\item \textbf{Handover and Multi-Connectivity:} Due to unreliability of individual mmWave links, dense deployments of small cells with fast handover protocols are critical in maintaining stable connections and avoiding TCP timeouts.
\item \textbf{\gls{cc} Algorithms:} With remote servers, we observe higher performance variations across different congestion control algorithms, while the difference is almost negligible with edge servers. Overall, BBR outperforms loss-based TCP
in terms of both rate and latency.

\begin{figure*}[t]
\begin{center}
\setlength\belowcaptionskip{-.45cm}
\includegraphics[width=.88\textwidth]{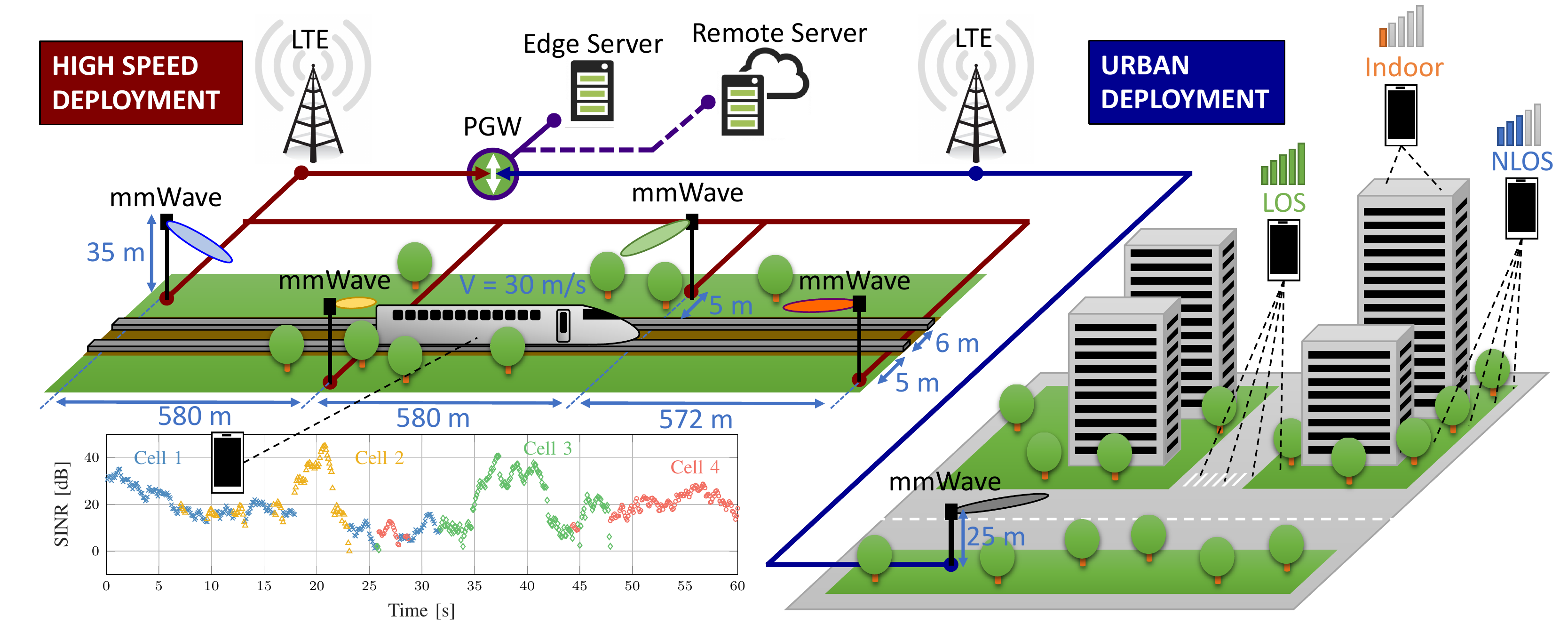}
\caption{High speed and urban deployment scenarios}
\label{fig:scenarios}
\end{center}
\end{figure*}

\item \textbf{TCP Packet Size:} We quantitatively compare the benefits of transmitting larger TCP packets in \gls{lte} versus mmWave networks, and show that, given the fluctuating Gbps data rates offered at mmWave frequencies, a larger packet size provides a faster growth of the congestion window and higher achievable rate.

\item \textbf{\gls{rlc} Buffer Size:} 
We analyze TCP performance over small and large buffers. 
While the TCP goodput degradation caused by buffer overflow in undersized buffers is difficult to mitigate, the problem of bufferbloating, i.e., large buffer occupancy leading to delays, can be approached by appropriately designing cross-layer algorithms~\cite{mengleiBufferbloat}. 
\end{enumerate}

The rest of the article is organized as follows. 
We first describe the scenarios of interest in Sec.~\ref{sec:scenarios}. Then, we list the main features of the \gls{cc} algorithms considered in this study in Sec.~\ref{sec:cc}. We report the main results and observations in Sec.~\ref{sec:results}, and draw our conclusions in Sec.~\ref{sec:conclusions}.

\section{5G Deployment Scenarios}
\label{sec:scenarios}
In order to assess how TCP will perform in mmWave cellular networks, we consider two of the most challenging scenarios among those specified by the 3GPP in~\cite{3GPP38913}, i.e., a high speed train and a dense urban scenario, represented in Fig.~\ref{fig:scenarios}. They were studied using the ns--3-based mmWave end-to-end simulation framework described in~\cite{mezzavilla2017end}, which models radio access, the core network, and the 3GPP channel for the mmWave band with spatial correlation in mobility scenarios. Moreover, the protocol stack simulated by~\cite{mezzavilla2017end} also features retransmissions at both the MAC layer, with \gls{harq}, and the \gls{rlc} layer, using the acknowledged mode option.

\textbf{High speed scenario:} In this scenario, shown on the left side of Fig.~\ref{fig:scenarios}, we test the performance of TCP over a channel that varies frequently in time and under realistic mobility conditions. Multiple \glspl{gnb} provide coverage to the railway, which is mostly \gls{los}: even if the current \gls{gnb} is blocked by obstacles placed between \glspl{gnb} 2 and 3, the \gls{ue} can quickly perform a handover to another \gls{los} \gls{gnb}. 
The \glspl{gnb} are at a height of 35 meters, with an intersite distance of 580 meters. The train moves at a speed of 108 km/h, and, as a result, the channel experienced by the \gls{ue} varies very quickly because of severe fading and the Doppler effect, and, on a longer time scale, due to obstacles, as shown in the \gls{sinr} plot of Fig.~\ref{fig:scenarios}. 
We use the channel tracking and mobility scheme described in~\cite{poleseHo}, which features fast and locally coordinated handovers for devices that are dual-connected to a mmWave \gls{gnb} and a sub-6 GHz \gls{gnb} (e.g., an \gls{lte} base station). 

\textbf{Dense urban scenario:} In this deployment, shown on the right side of Fig.~\ref{fig:scenarios}, we study the fairness of TCP flows over multiple \glspl{ue} with different channel conditions. A single mmWave \gls{gnb} placed at a height of 25 meters serves a group of ten users moving at walking speed. They are located in different positions, in order to account for a mixture of channel conditions: four \glspl{ue} are in \gls{los}, thus perceiving a very high \gls{sinr}, four are in \gls{nlos} and the last two are inside a building, so that the received power is additionally attenuated by the building penetration loss. 

For both scenarios we consider two deployments of the TCP server which acts as the endpoint of the connection. 
The first is a traditional setup in which the server is hosted in a remote data center, with a minimum \gls{rtt} in the order of 40~ms, accounting for the latencies of both the core network and the public internet. The second is a \gls{mec} scenario~\cite{mach2017mobile}, in which the server is located close to the \glspl{gnb} with smaller latency (of the order of 4~ms).

\section{TCP Congestion Control Protocols}
\label{sec:cc}
In this section, we will describe the congestion control protocols and the TCP performance enhancement techniques considered in this paper.

\subsection{TCP Congestion Control Algorithms}
We study four most commonly used \gls{cc} algorithms.

\textbf{TCP NewReno} has been the default algorithm for the majority of communication systems. In the congestion avoidance phase, the congestion window \texttt{cwnd} is updated after the reception of every ACK. The update is based on the \gls{aimd} design: \texttt{cwnd} is increased by summing a term $\alpha/$\texttt{cwnd} for each received ACK, and divided by a factor $\beta$ for each packet loss. For NewReno these parameters are fixed to $\alpha=1$ and $\beta=2$. 

\textbf{HighSpeed TCP} is designed for high \gls{bdp} networks, in which NewReno may exhibit a very slow growth of the congestion window. HighSpeed behaves the same as NewReno when the congestion window is small, but when it exceeds a predefined threshold the parameters $\alpha$, $\beta$ become functions of the congestion window, in order to maintain a large \texttt{cwnd}. Moreover, the window growth of NewReno and HighSpeed depends on the ACK reception rate, thus a shorter \gls{rtt} increases the ACK frequency and further speeds up the window growth. 

\textbf{TCP CUBIC}, instead, increases the congestion window over time, without considering the ACK reception rate but rather capturing the absolute time since the last packet loss and using a cubical increase function for \texttt{cwnd}. It has been designed to increase the ramp-up speed of each connection while maintaining fairness with other users.

\textbf{TCP BBR}, recently presented by Google \cite{cardwell2016bbr}, measures bottleneck bandwidth and round-trip propagation time, or BBR, to perform congestion control. It strives to match the sending rate to the estimated bottleneck bandwidth by pacing packets and setting the congestion window to \texttt{cwnd} gain $\times$ BDP, where the \texttt{cwnd} gain is a factor $(\leqslant 2)$ that is used to balance the effects of delayed, stretched and aggregated ACKs on bandwidth estimation. 

\subsection{TCP Performance Enhancement Techniques}
\label{sec:pet}
The performance of TCP has been the object of many studies over the last decades, and, besides new \gls{cc} algorithms, many other techniques have been proposed and deployed either at the endpoints of the connection (TCP sender and receiver) or inside the network. 

 In case of multiple packet losses, the TCP \gls{sack} option~\cite{RFC2018} allows the receiver to inform the sender which packets were received successfully, so that the sender can retransmit only those which were actually lost. This dramatically improves the efficiency of the TCP retransmission process.

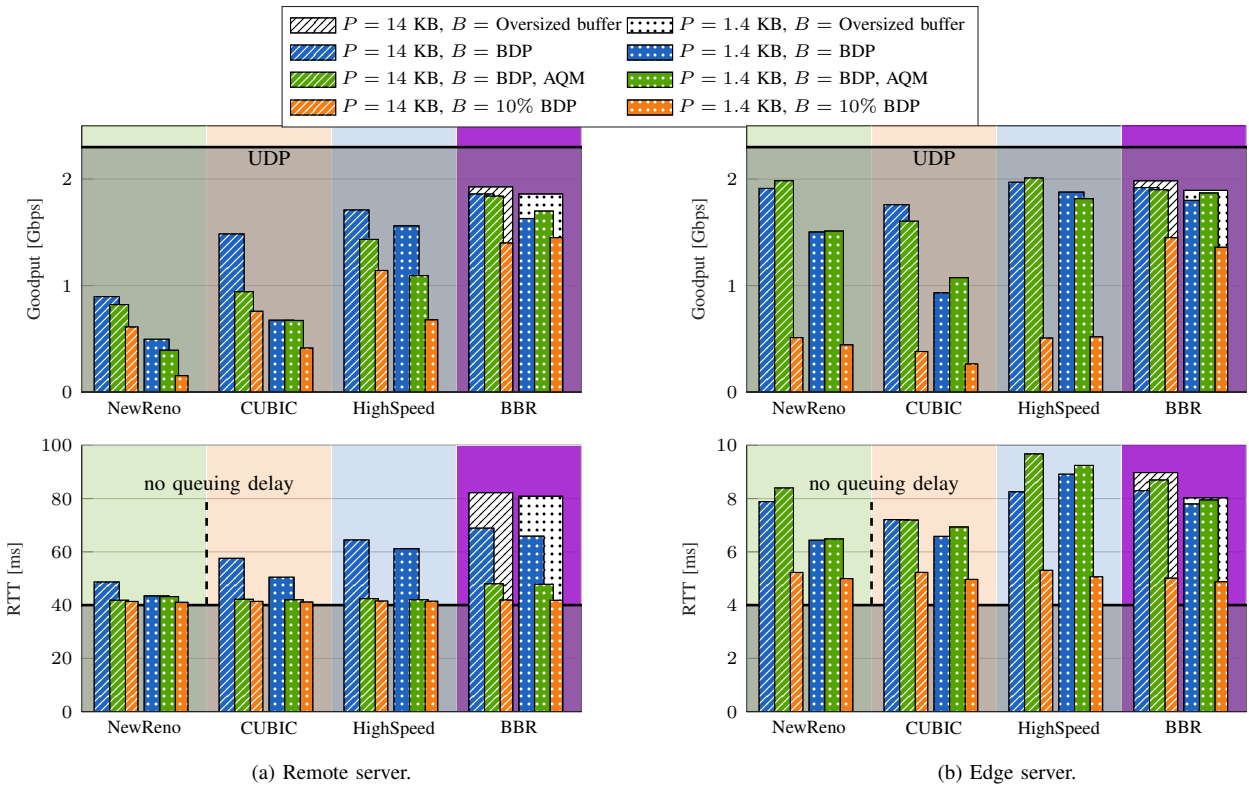
\begin{figure*}[t!]
  \centering
  \begin{subfigure}[t]{\columnwidth}
  \setlength\fwidth{0.75\columnwidth}
  \setlength\fheight{0.4\columnwidth}
%
%
\begin{tikzpicture}
\pgfplotsset{every tick label/.append style={font=\scriptsize}}
\pgfplotsset{scaled y ticks=false}
\definecolor{color0}{RGB}{83,164,0}
\definecolor{color1}{RGB}{255,255,255}
\definecolor{color2}{RGB}{31,100,191}
\definecolor{color3}{RGB}{242,125,20}

\definecolor{mycolor1}{RGB}{83,164,0}
\definecolor{mycolor2}{RGB}{150,0,200}
\definecolor{mycolor3}{RGB}{31,100,191}
\definecolor{mycolor4}{RGB}{242,125,20}

\begin{axis}[%
width=\fwidth,
height=\fheight,
at={(0in,1.2*\fheight)},
scale only axis,
xmin=0.5,
xmax=4.5,
xtick={1,2,3,4},
xticklabels={{NewReno},{CUBIC},{HighSpeed},{BBR}},
ymin=0,
ymax=2.5,
ymajorgrids=true,
ylabel style={font=\scriptsize\color{white!15!black}},
ylabel={Goodput [Gbps]},
axis background/.style={fill=white, draw=black},
title style={font=\bfseries},
axis x line*=bottom,
axis y line*=left,
legend style={font=\scriptsize, legend cell align=left, align=left, draw=white!15!black,at={(0.4,1.45)},anchor=north west},
legend columns=2,
]

\addplot[ybar, bar width=1, fill=mycolor1, opacity=0.2, draw=white, area legend,forget plot] table[row sep=crcr] {%
1 3\\
};

\addplot[ybar, bar width=1, fill=mycolor4, opacity=0.2, draw=white, area legend,forget plot] table[row sep=crcr] {%
2 3\\
};
\addplot[ybar, bar width=1, fill=mycolor3, opacity=0.2, draw=white, area legend,forget plot] table[row sep=crcr] {%
3 3\\
};
\addplot[ybar, bar width=1, fill=mycolor2, opacity=0.8, draw=white, area legend,forget plot] table[row sep=crcr] {%
4 3\\
};

\addplot[ybar, bar width=5, fill=gray, opacity=0.5, draw=black, area legend,forget plot] table[row sep=crcr] {%
2 2.3\\
};

\addplot[ybar, bar width=0.35, fill=color1, draw=black, area legend,
postaction={pattern=north east lines, pattern color=black}] table[row sep=crcr] {%
3.775 1.927\\
};
\addlegendentry{$P=14$~KB, $B=$ Oversized buffer}

\addplot[ybar, bar width=0.35, fill=color1, draw=black, area legend,
postaction={pattern=dots,pattern color=black}] table[row sep=crcr] {%
4.175 1.86\\
};
\addlegendentry{$P=1.4$~KB, $B=$ Oversized buffer}

\addplot[ybar, bar width=0.2, fill=color2, draw=black, area legend,
postaction={pattern=north east lines, pattern color=white}] table[row sep=crcr] {%
0.7	0.8968\\
1.7	1.4848\\
2.7	1.7095\\
3.7    1.86\\
};
\addplot[forget plot, color=white!15!black] table[row sep=crcr] {%
0.5	0\\
4.5	0\\
};
\addlegendentry{$P=14$~KB, $B=$ \gls{bdp}}

\addplot[ybar, bar width=0.2, fill=color2, draw=black, area legend,
postaction={pattern=dots,pattern color=white}] table[row sep=crcr] {%
1.1	0.4957\\
2.1	0.6752\\
3.1	1.5613\\
4.1    1.63\\
};
\addplot[forget plot, color=white!15!black] table[row sep=crcr] {%
0.5	0\\
4.5	0\\
};
\addlegendentry{$P=1.4$~KB, $B=$ \gls{bdp}}

\addplot[ybar, bar width=0.15, fill=color0, draw=black, area legend,
postaction={pattern=north east lines, pattern color=white}] table[row sep=crcr] {%
0.8	0.8215\\
1.8	0.9428\\
2.8	1.4332\\
3.8    1.84\\
};
\addplot[forget plot, color=white!15!black] table[row sep=crcr] {%
0.5	0\\
4.5	0\\
};
\addlegendentry{$P=14$~KB, $B=$ \gls{bdp}, AQM}

\addplot[ybar, bar width=0.15, fill=color0, draw=black, area legend,
postaction={pattern=dots,pattern color=white}] table[row sep=crcr] {%
1.2	0.3928\\
2.2	0.6727\\
3.2	1.0933\\
4.2   1.7\\
};
\addplot[forget plot, color=white!15!black] table[row sep=crcr] {%
0.5	0\\
4.5	0\\
};
\addlegendentry{$P=1.4$~KB, $B=$ \gls{bdp}, AQM}

\addplot[ybar, bar width=0.1, fill=color3, draw=black, area legend,
postaction={pattern=north east lines, pattern color=white}] table[row sep=crcr] {%
0.9	0.6114\\
1.9	0.7588\\
2.9	1.1418\\
3.9    1.4\\
};
\addplot[forget plot, color=white!15!black] table[row sep=crcr] {%
0.5	0\\
4.5	0\\
};
\addlegendentry{$P=14$~KB, $B=10\%$ \gls{bdp}}

\addplot[ybar, bar width=0.1, fill=color3, draw=black, area legend,
postaction={pattern=dots,pattern color=white}] table[row sep=crcr] {%
1.3	0.1544\\
2.3	0.4136\\
3.3	0.6788\\
4.3   1.45\\
};
\addplot[forget plot, color=white!15!black] table[row sep=crcr] {%
0.5	0\\
4.5	0\\
};
\addlegendentry{$P=1.4$~KB, $B=10\%$ \gls{bdp}}

\addplot[color=black, line width=1] coordinates {(-6,2.3) (6,2.3)};
\addplot[dashed, color=black, line width=1] coordinates {(1.5,40) (1.5,59)};
 \node[color=black, font=\footnotesize] at (2,2.2) {UDP};

\end{axis}

\begin{axis}[%
width=\fwidth,
height=\fheight,
at={(0in,0in)},
scale only axis,
xmin=0.5,
xmax=4.5,
xtick={1,2,3,4},
xticklabels={{NewReno},{CUBIC},{HighSpeed},{BBR}},
ymin=0,
ymax=100,
ymajorgrids=true,
ylabel style={font=\scriptsize\color{white!15!black}},
ylabel={RTT [ms]},
axis background/.style={fill=white, draw=black},
axis x line*=bottom,
axis y line*=left,
]

\addplot[ybar, bar width=1, fill=mycolor1, opacity=0.2, draw=white, area legend] table[row sep=crcr] {%
1 100\\
};
\addplot[ybar, bar width=1, fill=mycolor4, opacity=0.2, draw=white, area legend] table[row sep=crcr] {%
2 100\\
};
\addplot[ybar, bar width=1, fill=mycolor3, opacity=0.2, draw=white, area legend] table[row sep=crcr] {%
3 100\\
};
\addplot[ybar, bar width=1, fill=mycolor2, opacity=0.8, draw=white, area legend] table[row sep=crcr] {%
4 100\\
};

\addplot[ybar, bar width=5, fill=gray, opacity=0.5, draw=black, area legend,forget plot] table[row sep=crcr] {%
2 40\\
};
\addplot[color=black, line width=1] coordinates {(-6,40) (6,40)};
\addplot[dashed, color=black, line width=1] coordinates {(1.5,40) (1.5,80)};
 \node[color=black, font=\footnotesize] at (1.6,85) {no queuing delay};

\addplot[ybar, bar width=0.35, fill=color1, draw=black, area legend,
postaction={pattern=north east lines, pattern color=black}] table[row sep=crcr] {%
3.775 82.22\\
};

\addplot[ybar, bar width=0.35, fill=color1, draw=black, area legend,
postaction={pattern=dots,pattern color=black}] table[row sep=crcr] {%
4.175 80.874\\
};

\addplot[ybar, bar width=0.2, fill=color2, draw=black, area legend,
postaction={pattern=north east lines, pattern color=white}] table[row sep=crcr] {%
0.7	48.6706\\
1.7	57.5675\\
2.7	64.4569\\
3.7  68.9\\
};
\addplot[forget plot, color=white!15!black] table[row sep=crcr] {%
0.5	0\\
4.5	0\\
};

\addplot[ybar, bar width=0.2, fill=color2, draw=black, area legend,
postaction={pattern=dots,pattern color=white}] table[row sep=crcr] {%
1.1	43.46\\
2.1	50.4662\\
3.1	61.1986\\
4.1   65.9\\
};
\addplot[forget plot, color=white!15!black] table[row sep=crcr] {%
0.5	0\\
4.5	0\\
};

\addplot[ybar, bar width=0.15, fill=color0, draw=black, area legend,
postaction={pattern=north east lines, pattern color=white}] table[row sep=crcr] {%
0.8	41.8202\\
1.8	42.1776\\
2.8	42.4351\\
3.8   48\\
};
\addplot[forget plot, color=white!15!black] table[row sep=crcr] {%
0.5	0\\
4.5	0\\
};

\addplot[ybar, bar width=0.15, fill=color0, draw=black, area legend,
postaction={pattern=dots,pattern color=white}] table[row sep=crcr] {%
1.2	43.1986\\
2.2	41.9896\\
3.2	42.0189\\
4.2   47.8\\
};
\addplot[forget plot, color=white!15!black] table[row sep=crcr] {%
0.5	0\\
4.5	0\\
};

\addplot[ybar, bar width=0.1, fill=color3, draw=black, area legend,
postaction={pattern=north east lines, pattern color=white}] table[row sep=crcr] {%
0.9	41.3685\\
1.9	41.3911\\
2.9	41.5044\\
3.9   41.9\\
};
\addplot[forget plot, color=white!15!black] table[row sep=crcr] {%
0.5	0\\
4.5	0\\
};

\addplot[ybar, bar width=0.1, fill=color3, draw=black, area legend,
postaction={pattern=dots,pattern color=white}] table[row sep=crcr] {%
1.3	41.0427\\
2.3	41.1536\\
3.3	41.4399\\
4.3   41.8\\
};
\addplot[forget plot, color=white!15!black] table[row sep=crcr] {%
0.5	0\\
4.5	0\\
};

\end{axis}
\end{tikzpicture}%
\caption{Remote server.}
\label{fig:40ms}
\end{subfigure}
  \begin{subfigure}[t]{\columnwidth}
  \setlength\fwidth{0.75\columnwidth}
  \setlength\fheight{0.4\columnwidth}
%
%
\begin{tikzpicture}
\pgfplotsset{every tick label/.append style={font=\scriptsize}}
\pgfplotsset{scaled y ticks=false}
\definecolor{color0}{RGB}{83,164,0}
\definecolor{color1}{RGB}{255,255,255}

\definecolor{color2}{RGB}{31,100,191}
\definecolor{color3}{RGB}{242,125,20}

\definecolor{mycolor1}{RGB}{83,164,0}
\definecolor{mycolor2}{RGB}{150,0,200}
\definecolor{mycolor3}{RGB}{31,100,191}
\definecolor{mycolor4}{RGB}{242,125,20}

\begin{axis}[%
width=\fwidth,
height=\fheight,
at={(0in,1.2*\fheight)},
scale only axis,
xmin=0.5,
xmax=4.5,
xtick={1,2,3,4},
xticklabels={{NewReno},{CUBIC},{HighSpeed},{BBR}},
ymin=0,
ymax=2.5,
ymajorgrids=true,
ylabel style={font=\scriptsize\color{white!15!black}},
ylabel={Goodput [Gbps]},
axis background/.style={fill=white, draw=black},
title style={font=\bfseries},
axis x line*=bottom,
axis y line*=left,
legend style={font=\scriptsize, legend cell align=left, align=left, draw=white!15!black,at={(0.5,1.35)},anchor=north},
legend columns=2,
]

\addplot[ybar, bar width=1, fill=mycolor1, opacity=0.2, draw=white, area legend,forget plot] table[row sep=crcr] {%
1 3\\
};

\addplot[ybar, bar width=1, fill=mycolor4, opacity=0.2, draw=white, area legend,forget plot] table[row sep=crcr] {%
2 3\\
};
\addplot[ybar, bar width=1, fill=mycolor3, opacity=0.2, draw=white, area legend,forget plot] table[row sep=crcr] {%
3 3\\
};
\addplot[ybar, bar width=1, fill=mycolor2, opacity=0.8, draw=white, area legend,forget plot] table[row sep=crcr] {%
4 3\\
};

\addplot[ybar, bar width=5, fill=gray, opacity=0.5, draw=black, area legend,forget plot] table[row sep=crcr] {%
2 2.3\\
};

\addplot[color=black, line width=1] coordinates {(-6,2.3) (6,2.3)};
 \node[color=black, font=\footnotesize] at (2,2.2) {UDP};

\addplot[ybar, bar width=0.35, fill=color1, draw=black, area legend,
postaction={pattern=north east lines, pattern color=black}] table[row sep=crcr] {%
3.775 1.9839\\
};

\addplot[ybar, bar width=0.35, fill=color1, draw=black, area legend,
postaction={pattern=dots,pattern color=black}] table[row sep=crcr] {%
4.175 1.894\\
};
\addplot[ybar, bar width=0.2, fill=color2, draw=black, area legend,
postaction={pattern=north east lines, pattern color=white}] table[row sep=crcr] {%
0.7	1.9122\\
1.7	1.7591\\
2.7	1.9703\\
3.7    1.92\\
};
\addplot[forget plot, color=white!15!black] table[row sep=crcr] {%
0.5	0\\
4.5	0\\
};

\addplot[ybar, bar width=0.2, fill=color2, draw=black, area legend,
postaction={pattern=dots,pattern color=white}] table[row sep=crcr] {%
1.1	1.5035\\
2.1	0.9320\\
3.1	1.8779\\
4.1    1.8\\
};
\addplot[forget plot, color=white!15!black] table[row sep=crcr] {%
0.5	0\\
4.5	0\\
};

\addplot[ybar, bar width=0.15, fill=color0, draw=black, area legend,
postaction={pattern=north east lines, pattern color=white}] table[row sep=crcr] {%
0.8	1.9842\\
1.8	1.6045\\
2.8	2.0106\\
3.8    1.90\\
};
\addplot[forget plot, color=white!15!black] table[row sep=crcr] {%
0.5	0\\
4.5	0\\
};

\addplot[ybar, bar width=0.15, fill=color0, draw=black, area legend,
postaction={pattern=dots,pattern color=white}] table[row sep=crcr] {%
1.2	1.5125\\
2.2	1.0746\\
3.2	1.8164\\
4.2    1.87\\
};
\addplot[forget plot, color=white!15!black] table[row sep=crcr] {%
0.5	0\\
4.5	0\\
};

\addplot[ybar, bar width=0.1, fill=color3, draw=black, area legend,
postaction={pattern=north east lines, pattern color=white}] table[row sep=crcr] {%
0.9	0.5114\\
1.9	0.3810\\
2.9	0.5075\\
3.9   1.45\\
};
\addplot[forget plot, color=white!15!black] table[row sep=crcr] {%
0.5	0\\
4.5	0\\
};

\addplot[ybar, bar width=0.1, fill=color3, draw=black, area legend,
postaction={pattern=dots,pattern color=white}] table[row sep=crcr] {%
1.3	0.4441\\
2.3	0.2642\\
3.3	0.5194\\
4.3   1.36\\
};
\addplot[forget plot, color=white!15!black] table[row sep=crcr] {%
0.5	0\\
4.5	0\\
};

\end{axis}

\begin{axis}[%
width=\fwidth,
height=\fheight,
at={(0in,0in)},
scale only axis,
xmin=0.5,
xmax=4.5,
xtick={1,2,3,4},
xticklabels={{NewReno},{CUBIC},{HighSpeed},{BBR}},
ymin=0,
ymax=10,
ymajorgrids=true,
ylabel style={font=\scriptsize\color{white!15!black}},
ylabel={RTT [ms]},
axis background/.style={fill=white, draw=black},
axis x line*=bottom,
axis y line*=left,
]

\addplot[ybar, bar width=1, fill=mycolor1, opacity=0.2, draw=white, area legend] table[row sep=crcr] {%
1 100\\
};
\addplot[ybar, bar width=1, fill=mycolor4, opacity=0.2, draw=white, area legend] table[row sep=crcr] {%
2 100\\
};
\addplot[ybar, bar width=1, fill=mycolor3, opacity=0.2, draw=white, area legend] table[row sep=crcr] {%
3 100\\
};
\addplot[ybar, bar width=1, fill=mycolor2, opacity=0.8, draw=white, area legend] table[row sep=crcr] {%
4 100\\
};

\addplot[ybar, bar width=5, fill=gray, opacity=0.5, draw=black, area legend,forget plot] table[row sep=crcr] {%
2 4\\
};
\addplot[color=black, line width=1] coordinates {(-6,4) (6,4)};
\addplot[dashed, color=black, line width=1] coordinates {(1.5,4) (1.5,8)};
 \node[color=black, font=\footnotesize] at (1.6,8.5) {no queuing delay};
 
 \addplot[ybar, bar width=0.35, fill=color1, draw=black, area legend,
postaction={pattern=north east lines, pattern color=black}] table[row sep=crcr] {%
3.775 	8.977\\
};

\addplot[ybar, bar width=0.35, fill=color1, draw=black, area legend,
postaction={pattern=dots,pattern color=black}] table[row sep=crcr] {%
4.175	8.03\\
};

\addplot[ybar, bar width=0.2, fill=color2, draw=black, area legend,
postaction={pattern=north east lines, pattern color=white}] table[row sep=crcr] {%
0.7	7.8867\\
1.7	7.2101\\
2.7	8.2556\\
3.7   8.3\\
};
\addplot[forget plot, color=white!15!black] table[row sep=crcr] {%
0.5	0\\
4.5	0\\
};

\addplot[ybar, bar width=0.2, fill=color2, draw=black, area legend,
postaction={pattern=dots,pattern color=white}] table[row sep=crcr] {%
1.1	6.4365\\
2.1	6.5812\\
3.1	8.9156\\
4.1  7.8\\
};
\addplot[forget plot, color=white!15!black] table[row sep=crcr] {%
0.5	0\\
4.5	0\\
};

\addplot[ybar, bar width=0.15, fill=color0, draw=black, area legend,
postaction={pattern=north east lines, pattern color=white}] table[row sep=crcr] {%
0.8	8.3993\\
1.8	7.194\\
2.8	9.6767\\
3.8    8.7\\
};
\addplot[forget plot, color=white!15!black] table[row sep=crcr] {%
0.5	0\\
4.5	0\\
};

\addplot[ybar, bar width=0.15, fill=color0, draw=black, area legend,
postaction={pattern=dots,pattern color=white}] table[row sep=crcr] {%
1.2	6.4866\\
2.2	6.9359\\
3.2	9.2454\\
4.2   7.95\\
};
\addplot[forget plot, color=white!15!black] table[row sep=crcr] {%
0.5	0\\
4.5	0\\
};

\addplot[ybar, bar width=0.1, fill=color3, draw=black, area legend,
postaction={pattern=north east lines, pattern color=white}] table[row sep=crcr] {%
0.9	5.2256\\
1.9	5.2278\\
2.9	5.3054\\
3.9  5.01\\
};
\addplot[forget plot, color=white!15!black] table[row sep=crcr] {%
0.5	0\\
4.5	0\\
};

\addplot[ybar, bar width=0.1, fill=color3, draw=black, area legend,
postaction={pattern=dots,pattern color=white}] table[row sep=crcr] {%
1.3	4.9956\\
2.3	4.9646\\
3.3	5.0606\\
4.3  4.88\\
};
\addplot[forget plot, color=white!15!black] table[row sep=crcr] {%
0.5	0\\
4.5	0\\
};

\end{axis}
\end{tikzpicture}%
\caption{Edge server.}
\label{fig:4ms}
\end{subfigure}
  \setlength\belowcaptionskip{-.45cm}
\caption{Goodput and \gls{rtt} for the high speed train scenario, with the remote and the edge server for different combinations of the buffer size and the MSS.}
\label{fig:hst}
\end{figure*}

\gls{aqm} schemes~\cite{gong2014fighting}, instead, are deployed in network devices (e.g., routers, gateways, \glspl{gnb}), to control the behavior of their queues and buffers. The size of these buffers plays an important role in the end-to-end performance. 
If the buffer is too small, many packets may be dropped when the buffer is full, according to the drop tail policy. 
Conversely, if the buffer is too large, then the bufferbloat phenomenon occurs~\cite{gong2014fighting}.
\gls{aqm} techniques can be deployed at the buffers to drop packets before the queue is full, so that the TCP sender can proactively react to the congestion that could arise in the near future.

Finally, there are some techniques that are typically used in combination with wireless links. The first is the usage of link-layer retransmissions between the \gls{gnb} and the \gls{ue}, so that the losses on the channel are masked from TCP. This helps increase the goodput, however the end-to-end latency also increases, as shown in~\cite{polese2017tcp}. Another technique which is often used in wireless networks is proxying~\cite{liu2016improving}, i.e., the connection is split into two at some level in the mobile network (e.g., at the gateway with the internet, at the \gls{gnb}, etc), and different \gls{cc} techniques are deployed over the two parts of the connection.

\section{TCP Performance in the 3GPP Scenarios}
\label{sec:results}

In the following paragraphs we will report the performance of the TCP congestion control algorithms presented in Sec.~\ref{sec:cc} over the 5G mmWave deployment scenarios described in Sec.~\ref{sec:scenarios}, focusing on both goodput and latency. The results are averaged over multiple independent simulation runs, so that the confidence intervals are small (they are however not shown to make the figures easier to read). In all the simulations, we use full buffer traffic with the TCP \gls{sack} option and disable the TCP delayed ACK mechanism, thus each received packet will generate an ACK. The minimum retransmission timeout is set to 200 ms.

\vspace{-.2cm}
\subsection{High Speed Deployment Scenario}
\vspace{-.1cm}
In this scenario we compare different combinations of the \gls{rlc} buffer size $B$ and the \gls{mss} $P$  with a single TCP connection from the \gls{ue}. 
For both the remote and the edge server deployments the \gls{rlc} buffer is 10\% or 100\% of the \gls{bdp} computed considering the maximum achievable data rate (3 Gbit/s) and the minimum latency, i.e., $B$ equals 1.5 or 15~MB for the remote server deployment, and 0.15 or 1.5~MB for the edge server. We also consider two different \gls{mss}, i.e., a standard \gls{mss} of 1400 bytes (1.4 KB) and a large \gls{mss} of 14000 bytes (14 KB). The goodput of saturated UDP traffic is also provided as a reference for the maximum achievable rate, as shown in Fig.~\ref{fig:hst}.

Notice that, thanks to the mobility management scheme based on dual connectivity and fast secondary cell handover, and despite the high mobility of the scenario, we never observed a TCP connection reset due to an outage, i.e., even if the closest two base stations are blocked, the \gls{ue} is still capable of receiving signals from other nearby \glspl{gnb}. Therefore, even if blockage events are still possible, in a scenario with a dense deployment (according to 3GPP guidelines), it is possible to provide uninterrupted connectivity to the final user~\cite{polese2017mobility}.

In the following paragraphs we will provide insights on the effects of the different parameters on TCP performance over mmWave at high speed.

\begin{figure*}[t]
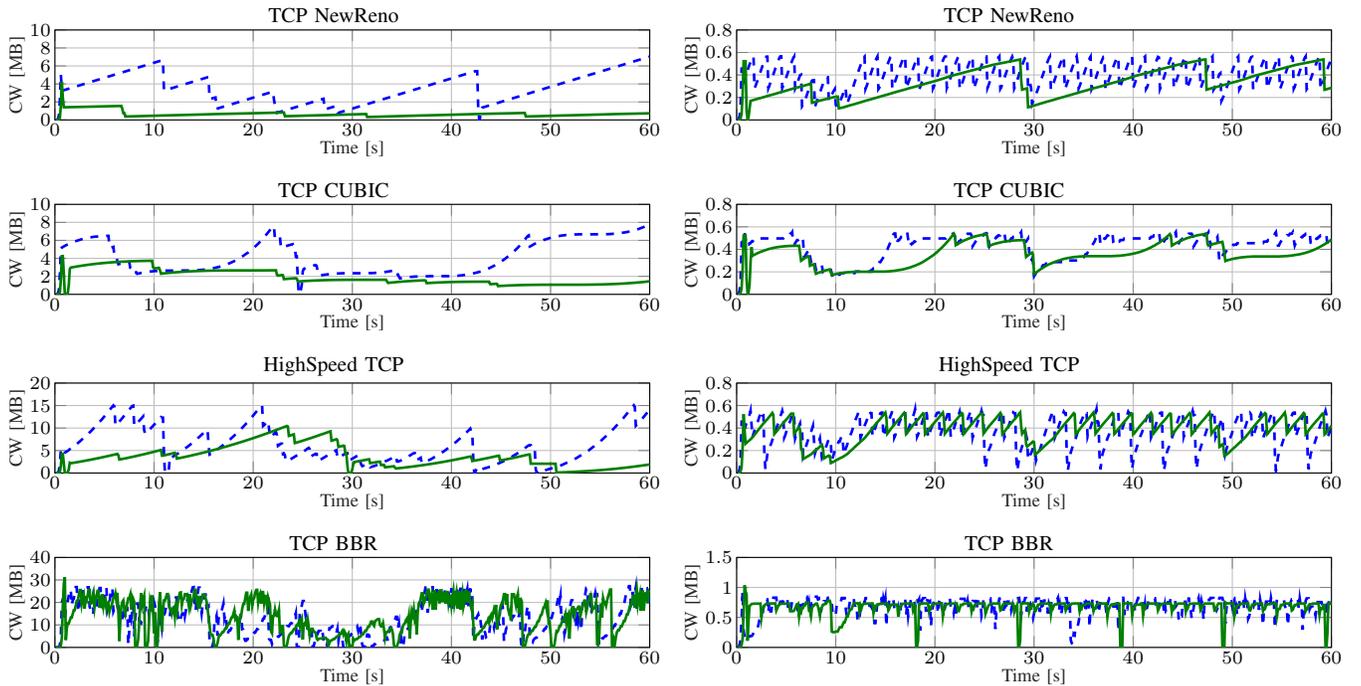

\centering
\begin{subfigure}[t]{\columnwidth}
  \setlength\fwidth{0.9\columnwidth}
  \setlength\fheight{0.85\columnwidth}
  \adjustbox{center}{\footnotesize TCP NewReno}\vspace*{-.1cm}
  \input{figure/cw_newReno.tex}
  \adjustbox{center}{\footnotesize TCP CUBIC}\vspace*{-.1cm}
  \input{figure/cw_cubic.tex}
  \adjustbox{center}{\footnotesize HighSpeed TCP}\vspace*{-.1cm}
  \input{figure/cw_hs.tex}
  \adjustbox{center}{\footnotesize TCP BBR}\vspace*{-.1cm}
  \input{figure/cw_bbr.tex}
  \caption{TCP congestion window with mmWave}
  \label{fig:cwEvoMmWave}
\end{subfigure}
\begin{subfigure}[t]{\columnwidth}
  \setlength\fwidth{0.9\columnwidth}
  \setlength\fheight{0.85\columnwidth}
  \adjustbox{center}{\footnotesize TCP NewReno}\vspace*{-.1cm}
  \input{figure/cw_newReno_lte.tex}
  \adjustbox{center}{\footnotesize TCP CUBIC}\vspace*{-.1cm}
  \input{figure/cw_cubic_lte.tex}
  \adjustbox{center}{\footnotesize HighSpeed TCP}\vspace*{-.1cm}
  \input{figure/cw_hs_lte.tex}
  \adjustbox{center}{\footnotesize TCP BBR}\vspace*{-.1cm}
  \input{figure/cw_bbr_lte.tex}
  \caption{TCP congestion window with \gls{lte}}
  \label{fig:cwEvoLte}
\end{subfigure}
\setlength\belowcaptionskip{-.3cm}
\caption{Congestion window evolution over time for different \gls{cc} algorithms. The scenario is configured with remote servers and small \gls{rlc} buffers}
  \label{fig:cwEvo}
\end{figure*}

\subsubsection{\textbf{Impact of the server deployment}}
Loss-based TCP benefits from the shorter control loop related to an edge server deployment, as shown by comparing Figs.~\ref{fig:40ms} and~\ref{fig:4ms}. With the latter, indeed, the differences between the maximum goodput of the loss-based TCP versions are less marked, since the faster reaction time makes up for the differences among them. 
Moreover, the \gls{rtt} difference between the large and the small \gls{rlc} buffer is lower in absolute terms (milliseconds with edge server versus tens of milliseconds with remote server), but the ratio is approximately the same. However, for CUBIC and HighSpeed with the smallest buffer configuration, the goodput is lower with the edge than with the remote server,  i.e., there is a 30\% loss with the smallest \gls{mss}, and of 50\% with the largest one. In this case, indeed, the buffer size is very small (i.e., $B=0.15$~MB), thus incurring buffer overflows\footnote{With large \gls{mss} just 11 packets are enough to cause a buffer overflow.}, which reduce the sending rate. 

\subsubsection{\textbf{Impact of the congestion control algorithm}}
The congestion control algorithm has a stronger impact in the remote server scenario. The best performance, in terms of goodput, is achieved by BBR with large buffer size, but it is still 400 Mbps lower than the maximum achievable rate. Moreover, as observed in \cite{zhang2016transport,polese2017tcp}, high goodput values also correspond to higher end-to-end latency. However, with small buffers, BBR produces the highest goodput (especially in the edge server scenario), with a latency comparable to loss-based TCP. BBR, indeed, regulates its sending rate to the estimated bandwidth and is not affected by packet loss, i.e., the congestion window dynamics of BBR, presented in Fig.~\ref{fig:cwEvoMmWave}, matches the SINR plot in Fig.~\ref{fig:scenarios}.

However, the loss-based versions of TCP cannot adjust their congestion window fast enough to adapt to the channel variations and perform worse than BBR, especially with small buffer, as seen in Fig.~\ref{fig:cwEvoMmWave}. Among them, TCP HighSpeed provides the highest goodput because of the aggressive window growth in the high \gls{bdp} region. TCP CUBIC performs better than NewReno in the remote server case, but worse in the edge server case. This is because CUBIC's window growth is not affected by the ACK rate, and therefore is more reliable over long \gls{rtt} links.

\subsubsection{\textbf{Impact of the \gls{mss}}}
\label{sec:mss}
The \gls{mss} does not affect the performance of BBR, which probes the bandwidth with a different mechanism, whereas, for loss-based TCP, the impact of the \gls{mss} on the goodput is remarkable.\footnote{Typically, TCP segments are mapped to multiple MAC/PHY data units, which complicates the dependence between a larger value of the TCP MSS and the correspondingly higher packet error probability over the wireless link. This non-trivial relationship, which would deserve a study by itself, has been properly captured in our numerical results.} The standard \gls{mss} of $P=$1.4 KB exhibits much worse performance compared to a larger \gls{mss} of $P=$14 KB. This happens because, in congestion avoidance, the congestion window increases by \gls{mss} bytes every \gls{rtt}, if all the packets are received correctly and delayed acknowledgment is not used, so the smaller the \gls{mss} the slower the window growth. 
Hence, the \gls{mss} dictates the congestion window's growth, which is particularly critical in mmWave networks for two main reasons: \emph{(i)} The mmWave peak capacity is at least one order of magnitude higher than in \gls{lte}, so that the congestion window will take a much longer time to reach the achievable link rate. In this case, we can gain in performance by simply using a larger \gls{mss}, as depicted in Fig. \ref{fig:hst}. \emph{(ii)} In addition, the channel fluctuations in the mmWave band will result in frequent quality drops, thus often requiring the congestion window to quickly ramp up to the link capacity to avoid underutilizing the channel.

\textbf{Large \gls{mss} -- mmWave vs. \gls{lte}}: 
Aimed at better illustrating why larger packets are particularly important in 5G mmWave networks, we also provide a performance comparison against \gls{lte} in the same scenario\footnote{For the \gls{lte} setup the small buffer represents 50\% of the \gls{bdp} (i.e., 0.08 and 0.2 MB for edge and remote server, respectively), because a 10\% \gls{bdp} buffer would be too small to protect from random fluctuations of the channel.}, and report in Table~\ref{table:lteMmWaveRatio} and Fig.~\ref{fig:cwEvo} detailed results focusing on the impact of the TCP \gls{mss} on the congestion window growth and, consequently, on the goodput of the system. Only a single user is placed in the high-speed train scenario, thus the drops in the congestion window are due to the worsening of the channel quality and not to contention with other flows.
Fig.~\ref{fig:cwEvo} shows that the loss-based TCP congestion window with a small \gls{mss} grows very slowly in congestion avoidance, and consequently loss-based TCP does not fully exploit the available bandwidth during the intervals in which the received signal has a very high \gls{sinr} (i.e., at $t=20$ s and $t=40$ s, as shown in Fig.~\ref{fig:scenarios}).
The large \gls{mss} helps speed up the congestion window's growth, which translates into higher goodput. Conversely, the goodput degradation associated with small packets is less relevant in \gls{lte} networks, given that the goodput is limited by the available bandwidth and not by the congestion window increase rate. These trends are reflected in Table~\ref{table:lteMmWaveRatio}. Among all loss-based TCP versions, only HighSpeed increases its congestion window fast enough even when transferring small packets. As a consequence, the goodput gain obtained with large \gls{mss} values is much smaller. 

Large packets introduce an additional benefit: due to (1) a reduced TCP/IP header overhead and (2) a reduced number of TCP ACKs, there will be more available downlink/uplink resources, resulting in higher goodput values. 

This solution may not be practical in an end-to-end network in which the \gls{mtu} is not entirely in control of the mobile network provider and is typically dictated by the adoption of Ethernet links (i.e., an \gls{mtu} of 1500 bytes).
By contrast, in a \gls{mec} scenario, in which the whole network is deployed by a single operator, it is possible to support a large \gls{mss} thanks to Ethernet~jumboframes~\cite{jumbo2}.

\begin{table}[t]
  \centering
  \renewcommand{\arraystretch}{1}
  \setlength\belowcaptionskip{-.7cm}
  \begin{tabular}{@{}llllll@{}}
  \toprule
  & & \multicolumn{2}{l}{Remote Server} & \multicolumn{2}{l}{Edge Server} \\
  \midrule
  & Buffer & BDP  & 10$\%$ BDP  & BDP & 10$\%$ BDP \\
  \midrule
  \multirow{2}{*}{TCP NewReno} & \gls{lte}   & 1.06 & 1.17 & 0.80 & 0.65\\
  & mmWave   & 1.81  & 3.96 & 1.27 & 1.15\\
  \midrule
  \multirow{2}{*}{TCP CUBIC} & \gls{lte}  & 1.06 & 1.15 & 1.03 & 0.89\\
  & mmWave & 2.2  & 1.83 & 1.89  & 1.44\\
  \midrule
  \multirow{2}{*}{HighSpeed TCP} & \gls{lte} & 1.08  & 0.9 & 0.94 & 0.95  \\
  & mmWave  & 1.09 & 1.69 & 1.05 & 0.98\\
    \midrule
  \multirow{2}{*}{TCP BBR} & \gls{lte}   & 1.00 & 0.96  & 1.02 & 0.82\\
  & mmWave & 1.14  & 0.97 & 1.06 & 1.06\\
  \bottomrule
  \end{tabular}
  \caption{Ratio between the goodput achieved with $P=14$ KB and with $P=1.4$ KB, for different configurations of the simulated scenario.}
  \label{table:lteMmWaveRatio}
\end{table}


\begin{figure*}[t!]
\centering
\setlength\belowcaptionskip{-.45cm}
  \setlength\fwidth{0.2\textwidth}
  \setlength\fheight{0.19\textwidth}
    \input{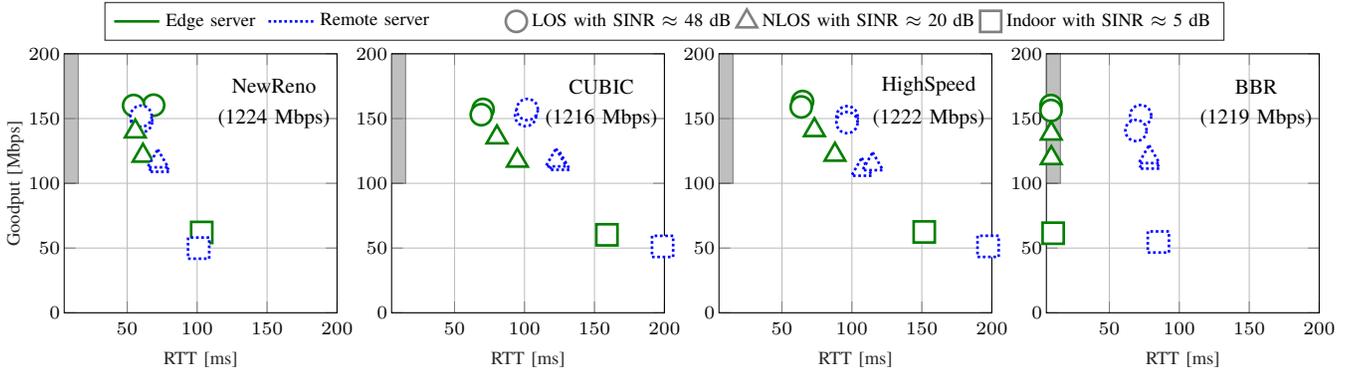}
\caption{Goodput vs \gls{rtt} for ten \glspl{ue} in the Urban Scenario, for different choices of the \gls{cc} algorithm.}
\label{fig:comp}
\end{figure*}

\subsubsection{\textbf{Impact of the buffer size and \gls{aqm}}}
The buffer size is also critical for the performance of TCP. As shown in Fig.~\ref{fig:hst}, large buffers generally yield higher goodput, because the probability of buffer overflow is smaller, and they offer a more effective protection against rapid and temporary variations of the mmWave channel quality. However, when a large buffer is coupled with loss-based TCP, the latency inevitably increases. Conversely, smaller buffers provide lower latency at the price of lower goodput.

For loss-based TCP, an intermediate solution is provided by applying \gls{aqm} to the largest buffer, especially in the remote server scenario. \gls{codel} is used as the default \gls{aqm} in our simulation because of its simple configuration. It controls the buffering latency by dropping packets when the minimum queuing delay within an interval, starting from 100 ms, is larger than 5 ms. \gls{codel} is picked as an example to show the trade-off between latency and goodput by using AQM. Our goal for this paper is not to select the best \gls{aqm} scheme or optimize \gls{aqm}, which in itself is a very interesting topic, and could be considered for future research. As shown in Fig.~\ref{fig:40ms}, the goodput with the \gls{aqm} option is larger than that with the smallest buffer, and in some cases (i.e., for the smallest packet size) is comparable to that of the \gls{bdp} buffer without \gls{aqm}, which in general yields the highest goodput. However, the latency is equivalent to the one associated with the small buffer, which is the lowest. In the edge server scenario the TCP control loop is short (the \gls{rtt} is 4 ms) and the reaction to congestion is quick. Hence, its performance is indeed equivalent to having \gls{bdp} buffers without \gls{aqm}.


BBR tries to solve this problem without modifying the buffers in the routers by maintaining a congestion window equal to twice the \gls{bdp} regardless of packet loss, as shown by Fig.~\ref{fig:cwEvo}. As a consequence, latency is only doubled in large buffers, and the goodput is slightly reduced in small buffers. These behaviors are also observed in the oversized and 10$\%$ \gls{bdp} buffer cases of Fig.~\ref{fig:hst}. 

%

\subsection{Urban Deployment Scenario}
\label{sec:urban}
In this scenario we consider ten \glspl{ue} attached to a single mmWave \gls{gnb}. In particular, we position four \glspl{ue} in \gls{los} conditions, four in \gls{nlos} and two inside a building. The average SINR for each channel condition is provided in Fig.~\ref{fig:comp}. Notice that, with low blockage density and walking speed, the channel condition is relatively stable over time. For each \gls{ue} pair one is connected to an edge server, and the other is connected to a remote server. In this way, it is possible to test the performance of TCP over a mixture of different conditions. The \gls{gnb} uses a \gls{rr} scheduler, so that the resource management at the base station does not have an impact on the fairness among different flows. All the \glspl{ue} use the same TCP version. We consider a standard \gls{mss} of 1400 bytes and an \gls{rlc} buffer size of 1.5 MB for each \gls{ue}.


Fig.~\ref{fig:comp} shows the average cell goodput (labeled in parentheses) and the goodput-latency trade-off for each type of user, separately, and for each \gls{cc} algorithm, in order to evaluate the fairness and the overall performance of different TCP versions with respect to different user channel conditions.

\begin{table*}[t]
  \centering
  \setlength\belowcaptionskip{-.45cm}
  \renewcommand{\arraystretch}{1}
  \begin{tabular}{@{}llllll@{}}
  \toprule
  & Loss-based & MSS impacts goodput & Summary & Considerations over 5G\\
  \midrule
  {TCP NewReno} & yes & yes & remote server: lowest goodput  &need to move servers to the edge\\
  \midrule
  {TCP CUBIC} & yes & yes & edge server: lowest goodput & need to increase MSS\\
  \midrule
  {HighSpeed TCP} & yes & only remote server & big buffer: high goodput and high latency & need to mitigate latency with AQM \\
    \midrule
   \multirow{2}{*}{TCP BBR} & \multirow{2}{*}{no} & \multirow{2}{*}{no} & big buffer: high goodput and high latency & small buffer is preferred \\ & & & \emph{small buffer: small rate reduction and low latency}   & performs well over mixed UE conditions\\
  \bottomrule
  \end{tabular}
  \caption{Results of the \gls{cc} algorithms over 5G deployments}
  \label{table:summarize}
\end{table*}

All \gls{cc} algorithms achieve the same average cell goodput, and similar goodput per \gls{ue}. However, the \gls{rtt} varies a lot among the \gls{cc} algorithms. The reason is that all \glspl{ue} use the same buffer size regardless of their channel conditions and network latency. As a consequence, the \gls{rlc} buffer size may be large for some \glspl{ue}, such as those at the edge. Therefore, the \gls{cc} algorithms that adopt a more aggressive window growth policy, such as CUBIC and HighSpeed, yield  much higher latency. 
For the loss-based TCP, \gls{nlos} and indoor \glspl{ue} suffer from a higher latency: given the same buffer size and backhaul data rate, a reduced bottleneck bandwidth results into an increased queueing delay in the buffers, until TCP settles to a steady state phase. BBR, instead, limits the congestion window to twice the estimated \gls{bdp}, and results in a maximum latency of 2 $\times$ minimum \gls{rtt}. We also draw a gray area in the plot representing a typical 5G application requirement, i.e., goodput greater than 100 Mbps and delay lower than 10 ms. In this scenario, among all \gls{cc} algorithms, only BBR meets this requirement for the \glspl{ue} connected to an edge server, and only under good channel conditions.

\vspace{-.2cm}
\section{Conclusions}
\vspace{-.1cm}
\label{sec:conclusions}
The massive but intermittent capacity available at mmWave frequencies introduces new challenges for all layers of the protocol stack, including TCP, the most widely used transport protocol. The interplay between congestion control algorithms and mmWave channel quality fluctuations makes the topic particularly complex, and represents the key driver behind this work. We have carried out a thorough simulation campaign, based on ns-3, across 3GPP-inspired scenarios, whose results are summarized in Table~\ref{table:summarize}. The main findings and some relevant research questions are listed as follows: (1) TCP benefits from a shorter control loop, where the server is placed at the cellular network edge and can react faster to link impairments. \emph{Should we (re)consider splitting TCP at some point?} (2) Moreover, when the \gls{rtt} is high, loss-based TCP underutilizes the mmWave capacity, while those based on congestion (e.g., BBR) show an improved performance by estimating the bandwidth of mmWave links. This means that new approaches based on \textit{more refined abstractions of the end-to-end network} can be studied for highly-variable and high-data-rate mmWave links. (3) Multi-connectivity and smart handovers, supported by advanced beamtracking and beamswitching techniques, will result in robust TCP connections. \emph{How densely should we deploy mmWave cells? 
How to support backhaul for densely deployed mmWave cells?}
(4) We show very clearly how loss-based TCP over mmWave bands can greatly benefit from using larger datagrams. \emph{Has the time come to break the legacy \gls{mtu} value}, by natively supporting larger packets in a wider set of networks? (5) Finally, it is well known that buffer size must scale proportionally to \gls{bdp} to achieve maximum TCP goodput. However, it is very challenging to properly dimension the buffers for mmWave links, given the rapid bandwidth variations between \gls{los} and \gls{nlos} conditions, and to protect from link losses without introducing bufferbloat. 
Given the low latency requirement and massive available bandwidth, \emph{is it beneficial to trade bandwidth for lower latency,} for example by running BBR over small \gls{rlc} buffer configurations?

We believe that these insights will stimulate further exploration of this important topic, which is essential to fully exploit the true potential of mmWave communications. Moreover, the observations provided by this initial simulation-based study can be used to guide the design of experimental activities, which are necessary to further validate the challenges that mmWave links pose to TCP, and to test novel techniques to improve the end-to-end user experience in mmWave cellular networks.


\bibliographystyle{IEEEtran}
\bibliography{bibliography.bib}

\newcommand{\SortNoop}[1]{}
\begin{thebibliography}{10}
\providecommand{\url}[1]{#1}
\csname url@samestyle\endcsname
\providecommand{\newblock}{\relax}
\providecommand{\bibinfo}[2]{#2}
\providecommand{\BIBentrySTDinterwordspacing}{\spaceskip=0pt\relax}
\providecommand{\BIBentryALTinterwordstretchfactor}{4}
\providecommand{\BIBentryALTinterwordspacing}{\spaceskip=\fontdimen2\font plus
\BIBentryALTinterwordstretchfactor\fontdimen3\font minus
  \fontdimen4\font\relax}
\providecommand{\BIBforeignlanguage}[2]{{%
\expandafter\ifx\csname l@#1\endcsname\relax
\typeout{** WARNING: IEEEtran.bst: No hyphenation pattern has been}%
\typeout{** loaded for the language `#1'. Using the pattern for}%
\typeout{** the default language instead.}%
\else
\language=\csname l@#1\endcsname
\fi
#2}}
\providecommand{\BIBdecl}{\relax}
\BIBdecl

\bibitem{Postel:81}
J.~Postel, ``Transmission control protocol,'' RFC 793, Sep. 1981.

\bibitem{liu2016improving}
K.~Liu and J.~Y. Lee, ``{On Improving TCP Performance over Mobile Data
  Networks},'' \emph{IEEE Trans. Mobile Comput.}, vol.~15, no.~10, pp.
  2522--2536, Oct. 2016.

\bibitem{xiao2017millimeter}
M.~Xiao, S.~Mumtaz, Y.~Huang, L.~Dai, Y.~Li, M.~Matthaiou, G.~K. Karagiannidis,
  E.~Björnson, K.~Yang, C.~L. I, and A.~Ghosh, ``{Millimeter Wave
  Communications for Future Mobile Networks},'' \emph{IEEE J. on Sel. Areas
  Commun.}, vol.~35, no.~9, pp. 1909--1935, Sept 2017.

\bibitem{zhang2016transport}
M.~Zhang, M.~Mezzavilla, R.~Ford, S.~Rangan, S.~Panwar, E.~Mellios, D.~Kong,
  A.~Nix, and M.~Zorzi, ``{Transport layer performance in 5G mmWave
  cellular},'' in \emph{IEEE Conference on Computer Communications Workshops
  (INFOCOM WKSHPS)}, April 2016, pp. 730--735.

\bibitem{polese2017tcp}
M.~Polese, R.~Jana, and M.~Zorzi, ``{TCP and MP-TCP in 5G mmWave Networks},''
  \emph{IEEE Internet Computing}, vol.~21, no.~5, pp. 12--19, Sept 2017.

\bibitem{mengleiBufferbloat}
M.~Zhang, M.~Mezzavilla, J.~Zhu, S.~Rangan, and S.~Panwar, ``{TCP dynamics over
  mmwave links},'' in \emph{IEEE 18th International Workshop on Signal
  Processing Advances in Wireless Communications (SPAWC)}, July 2017.

\bibitem{3GPP38913}
{3GPP}, ``{TR 38.913, Study on Scenarios and Requirements for Next Generation
  Access Technologies, V14.1.0},'' 2017.

\bibitem{mezzavilla2017end}
M.~Mezzavilla, M.~Zhang, M.~Polese, R.~Ford, S.~Dutta, S.~Rangan, and M.~Zorzi,
  ``{End-to-End Simulation of 5G mmWave Networks},'' \emph{IEEE Commun. Surveys
  Tuts.}, vol.~20, no.~3, pp. 2237--2263, third quarter 2018.

\bibitem{poleseHo}
M.~Polese, M.~Giordani, M.~Mezzavilla, S.~Rangan, and M.~Zorzi, ``{Improved
  Handover Through Dual Connectivity in 5G mmWave Mobile Networks},''
  \emph{IEEE J. Sel. Areas Commun.}, vol.~35, no.~9, pp. 2069--2084, Sept 2017.

\bibitem{mach2017mobile}
P.~Mach and Z.~Becvar, ``Mobile edge computing: A survey on architecture and
  computation offloading,'' \emph{IEEE Commun. Surveys Tuts.}, vol.~19, no.~3,
  pp. 1628--1656, third quarter 2017.

\bibitem{cardwell2016bbr}
N.~Cardwell, Y.~Cheng, C.~S. Gunn, S.~H. Yeganeh, and V.~Jacobson, ``{BBR}:
  Congestion-based congestion control,'' \emph{Queue}, vol.~14, no.~5, p.~50,
  2016.

\bibitem{RFC2018}
M.~Mathis, J.~Mahdavi, S.~Floyd, and A.~Romanow, ``{TCP Selective
  Acknowledgment Options},'' Internet Requests for Comments, RFC 2018, Oct.
  1996.

\bibitem{gong2014fighting}
Y.~Gong, D.~Rossi, C.~Testa, S.~Valenti, and M.~D. T{\"a}ht, ``{Fighting the
  bufferbloat: on the coexistence of AQM and low priority congestion
  control},'' \emph{Computer Networks}, vol.~65, pp. 255--267, June 2014.

\bibitem{polese2017mobility}
M.~Polese, M.~Mezzavilla, S.~Rangan, and M.~Zorzi, ``{Mobility Management for
  TCP in mmWave Networks},'' in \emph{Proceedings of the 1st ACM Workshop on
  Millimeter-Wave Networks and Sensing Systems 2017}, ser. mmNets '17, 2017,
  pp. 11--16.

\bibitem{jumbo2}
M.~Mezzavilla, D.~Chiarotto, D.~Corujo, M.~Wetterwald, and M.~Zorzi,
  ``{Evaluation of Jumboframes feasibility in LTE access networks},'' in
  \emph{IEEE International Conference on Communications (ICC)}, June 2013, pp.
  5964--5968.

\end{thebibliography}
\vspace*{-1cm}
\begin{IEEEbiographynophoto}{Menglei Zhang}
received the B.S. degree in electrical engineering from Nanjing University of Science and Technolodge, Nanjing, China, in 2013, and the M.S. degree in electrical engineering in 2015 from New York University Tandon School of Engineering, Brooklyn, NY, USA, where he is currently working toward the Ph.D. degree in electrical engineering with Prof. Rangan. His research interests include wireless communications, channel modeling, congestion control, and system level simulation.
\end{IEEEbiographynophoto}
\vspace*{-1cm}
\begin{IEEEbiographynophoto}{Michele Polese}
[S'17] received his B.Sc. (2014) and M.Sc. (2016) in Telecommunication Engineering from the University of Padova, Italy. Since October 2016 he has been a Ph.D. student at the Department of Information Engineering of the University of Padova. He visited New York University (NYU) and AT\&T Labs in Bedminster, NJ. His research interests focus on the analysis and development of protocols and architectures for 5G mmWave networks.
\end{IEEEbiographynophoto}
\vspace*{-1cm}
\begin{IEEEbiographynophoto}{Marco Mezzavilla}
is a research scientist at the NYU Tandon School of Engineering. He received his B.Sc. (2007), M.Sc. (2010) in telecommunications engineering, and Ph.D. (2013) in information engineering from the University of Padova, Italy. He is serving as reviewer for IEEE conferences, journals, and magazines. His research interests include design and validation of communication protocols and applications of 4G/5G wireless technologies, multimedia traffic optimization, radio resource management, spectrum sharing, convex optimization, cognitive networks, and experimental analysis.
\end{IEEEbiographynophoto}
\vspace*{-1cm}
\begin{IEEEbiographynophoto}{Jing Zhu}
[M'04-SM'12] received B.S. and M.S. degrees from Tsinghua University,
China in 1999 and 2001 respectively, and a Ph.D. in 2004 from University of Washington, Seattle, all in
electrical engineering. He is currently a Principal Engineer at Intel Corporation. His main research
interests are system design, performance optimization, and applications for heterogeneous wireless
networks, including 4G/5G cellular systems, high-density wireless LANs, and mobile ad hoc networks.
\end{IEEEbiographynophoto}
\vspace*{-1cm}
\begin{IEEEbiographynophoto}{Sundeep Rangan}
[F'15] is an associate professor of electrical and computer engineering at NYU and Associate 
Director of NYU WIRELESS. He received his Ph.D. from the University of California, Berkeley in electrical engineering. In 2000, he co-founded (with four others) Flarion Technologies, a spinoff of Bell Labs that developed Flash OFDM, the first cellular OFDM data system. It was acquired by Qualcomm in 2006, where he was a director of engineering prior to joining NYU in 2010.
\end{IEEEbiographynophoto}
\vspace*{-1cm}
\begin{IEEEbiographynophoto}{Shivendra Panwar}
[S'82-M'85-SM'00-F'11] is a Professor of Electrical and Computer Engineering at New York University. He is the Director of the New York State Center for Advanced Technology in Telecommunications (CATT), a member of NYU WIRELESS, and the Faculty Director of the NY City Media Lab. He was co-awarded the IEEE Communication Society’s Leonard G. Abraham Prize, and a co-author of the IEEE Multimedia Communications and ICC Best Papers for 2011 and 2016, respectively.
\end{IEEEbiographynophoto}
\vspace*{-1cm}
\begin{IEEEbiographynophoto}{Michele Zorzi}
[F'07] is with the Information Engineering Department of the University of Padova. His present research interests focus on various aspects of wireless communications. He was Editor-in-Chief of IEEE Wireless Communications from 2003 to 2005, IEEE Transactions on Communications from 2008 to 2011, and, at present, IEEE Transactions on Cognitive Communications and Networking. He served as a Member-at-Large of the ComSoc Board of Governors from 2009 to 2011, and as Director of Education and Training from 2014 to 2015.
\end{IEEEbiographynophoto}

\end{document}